\documentclass{ifacconf}
\usepackage{amsmath}
\usepackage{savesym}
\savesymbol{AND}
\savesymbol{OR}
\savesymbol{NOT}
\savesymbol{TO}
\savesymbol{COMMENT}
\savesymbol{BODY}
\savesymbol{IF}
\savesymbol{ELSE}
\savesymbol{ELSIF}
\savesymbol{FOR}
\savesymbol{WHILE}
\usepackage{algorithmic}
\usepackage{algorithm}
\usepackage{graphicx}
\usepackage{textcomp}
\usepackage{xcolor}
\usepackage{adjustbox}
\usepackage{graphicx}
\usepackage{amssymb}
\usepackage{lineno}
\usepackage{mathtools}
\usepackage{mathrsfs}
\usepackage{siunitx}
\usepackage{natbib} 
\usepackage{booktabs}   

\begin{document}
\begin{frontmatter}

\title{An incremental scenario approach for building energy management with uncertain occupancy \thanksref{footnoteinfo}} 

\thanks[footnoteinfo]{Support by the Pitch-In Project SMARTDISC is gratefully acknowledged. The work of A. Karshenas has been supported by a EUROP Fellowship, Department of Engineering Science, University of Oxford.}

\author[First]{Arman Karshenas} 
\author[First]{Kostas Margellos} 
\author[Second]{Simone Garatti}

\address[First]{Department of Engineering Science, University of Oxford, Oxford, UK (e-mail: arman.karshenasnajafabadi@balliol.ox.ac.uk).}
\address[Second]{Dipartimento di Elettronica e Informazione, 
   Politecnico di Milano, Milano, Italy, (e-mail: simone.garatti@polimi.it)}

\begin{abstract}                
We deal with the problem of energy management in buildings subject to uncertain occupancy. To this end, we formulate this as a finite horizon optimization program and optimize with respect to the windows' blinds position, radiator and cooling flux. Aiming at a schedule which is robust with respect to 
uncertain occupancy levels while avoiding imposing arbitrary assumptions on the underlying probability distribution of the uncertainty, we follow a data driven paradigm. In particular, we apply an incremental scenario approach methodology that has been recently proposed in the literature to our energy management formulation. To demonstrate the efficacy of the proposed implementation we provide a detailed numerical analysis on a stylized building and compare it with respect to a deterministic design and the standard scenario approach typically encountered in the literature. We show that our schedule is not agnostic with respect to uncertainty as deterministic approaches, while it requires fewer scenarios with respect to the standard scenario approach, thus resulting in a less conservative performance. 
\end{abstract}


\end{frontmatter}

\section{Introduction}
\label{intro}
The overall household energy use has increased noticeably over the last decade. As an example, in the United Kingdom, household consumption accounts for $50.2 \%$ of the total energy consumed \cite{department1}, \cite{department2}, while household and domestic heating/cooling is also responsible for over a quarter of the total emission levels \cite{department1}. At the same time there have been several advancements in instrumentation, control and efficiency of actuation, which have in turn boosted research towards optimal energy management within buildings.

To this end, optimization based control in building heating and cooling has been extensively studied both from a centralized perspective \cite{Frauke1}, \cite{ETH2},  \cite{ETH1}, \cite{MPC1}, \cite{swissoffice}, \cite{MPCbuilding}, and by means of a distributed architecture in multi-building settings \cite{Prandini1}, \cite{Prandini3}. 
In the aforementioned references, main concern has been the development of energy models for buildings suitable for being integrated within an optimization context, while occupancy has been considered to be deterministic. In the presence of uncertainty the attempts most closely related to our work have been proposed in \cite{Frauke2}, \cite{Prandini2}. In the former a robust approach is proposed (see also  \cite{robust1}, \cite{robust2} for conceptually similar robust considerations), with uncertainty assumed to be confined in sets with given geometry ignoring the underlying distribution of the uncertainty; the latter, may lead to conservative behaviour and deteriorate performance. In \cite{Prandini2}, a data driven approach is proposed using the so called scenario approach theory for convex optimization \cite{Nsufficient}, \cite{scenario2}, \cite{MPCcontrol}, \cite{simoneMPC}, thus alleviating the need of imposing certain assumptions on the underlying probability distribution of the uncertainty or the geometry of its support. 

In this paper, we follow the data driven route and represent occupancy by means of scenarios. We then determine a sequence of blind positions and heater flux settings, which remains feasible when a new realization of the uncertainty/occupancy is encountered.
We extend the developments in \cite{Prandini2} using an incremental scenario approach algorithm that has been recently proposed in the literature \cite{incscn}, and is motivated by the developments of \cite{waitjudge}. It leverages on the fact that the sample size proposed by the standard scenario approach is tight only for a certain class of programs (termed fully supported \cite{scenario2}), and for problems typically encountered in applications may result in conservative performance.
The incremental scheme gradually introduces additional scenarios to a given initial set of scenarios and terminates when the empirical estimate of the support constraints, a notion at the core of the scenario approach theory \cite{introbook}, is no greater than the current algorithm iteration. It is guaranteed that the incremental scenario approach leads to a performance no worse than the standard scenario approach, thus reducing both the level of conservatism of the resulting solution and the sample size required. The latter is of significant importance if sampling is expensive or if only a limited amount of historical data is available. We apply the incremental scenario approach on an energy management problem for a three-zone building, and compare it against a deterministic approach and the standard scenario approach implementation.

The energy management problem under consideration is introduced in Section \ref{secII}. Background on the scenario approach and the adopted incremental algorithm is presented in Section \ref{algorithm}. Section \ref{simulation} provides a detailed simulation based study, while Section \ref{sec:concl} concludes the paper and provides directions for future work. 

\section{Problem Statement} \label{secII}
\subsection{Description and mathematical modelling}
\label{Desandmodel}
\begin{figure}[t]
	\centering
\includegraphics[width=8.4cm]{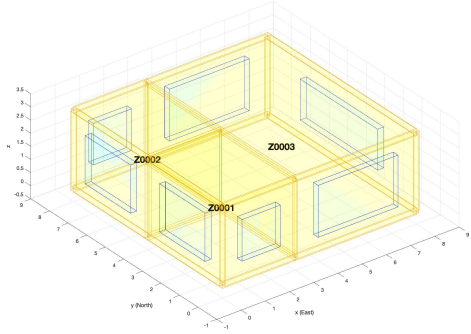}
	\caption{3D plot of the considered building model, generated using the MATLAB Toolbox of  \cite{ETH1}.}
	\label{buildingfloorand3D}
\end{figure}

We consider the heating and cooling problem for a symmetric building with two bedrooms and a living room labelled by Z0001, Z0002 and Z0003, respectively, in Figure \ref{buildingfloorand3D}. We selected a symmetric case here as it allows verifying the validity of the obtained results, with temperature profiles being identical in the symmetric zones (see upper left panel of Figure \ref{tempprofile}).

We consider a finite-horizon decision problem, where we aim at identifying an optimal sequence of actions for the following quantities, that are considered as inputs. These involve
\begin{enumerate}
	\item Windows' blinds position ($u_1$);
	\item Positive (heating) heat flux in $W/m^2$ by radiators or any external heating sources ($u_2$);
	\item Negative (cooling) heat flux in $W/m^2$ by cooling pipes or any external cooling sources ($u_3$). 
\end{enumerate}

For our analysis we consider ambient temperature and solar radiation to be deterministic, while building's
occupancy is not fixed and is considered to be uncertain. Under this setting we aim at identifying a sequence of the aforementioned inputs that are (probabilistically) robust with respect to uncertainty in occupancy levels. In particular, we model uncertainty as a heat flux in $W/m^2$ generated by an uncertain level of occupancy.

To model the heat transfer dynamics we employed the BRCM toolbox developed by \cite{ETH1}. The internal heat transfer between zones is governed by a linear time-invariant model while the external heat fluxes into the building introduce bilinear terms according to the nature of actuation. The modelling framework of \cite{ETH1} considers that i) the air volume of each zone has uniform temperature; ii) 
temperature within building elements varies only along the direction of the normal surface; iii)
there is no conductive heat transfer between different building elements; iv) the temperature within a layer of a building element is constant; v) all model parameters are constant over time; vi) long-wave, thermal radiation is considered in a combined convective heat transfer coefficient.
Under these assumptions, 
the overall dynamical model of heat transfer is then given by the following discrete time, bilinear system; more details about the structure of the model and the energy exchange can be found in \cite{ETH2},  \cite{ETH1}.

\begin{align}
	x_{k+1} = A x_k &+ B_u u_k + B_\delta \delta_k \nonumber \\
	& + \sum_{i=1}^n \big( B_{u\delta} \delta_k + B_{xu} x_k \big) u_{i,k},
	\label{bilineardiscrete}
	\end{align}
where for each time instance $k$, $x_k$ denotes the system state which contains the temperature of each zone and wall layer; $u_k$ contains the actuation inputs, and $\delta_k$ the uncertain occupancy level. The total number of inputs is denoted by $n$. All matrices are of appropriate dimensions. 

Our goals is to determine an optimal sequence $u$ by integrating the dynamical system in \eqref{bilineardiscrete} within a convex optimization context. To this end, the bilinear terms impose a challenge, as they would render the problem non-convex. To alleviate this we impose the following simplifying assumptions. 
\begin{enumerate}
	\item The ambient air temperature during the summer is assumed to be constant at $\ang{35}C$;
	\item The global radiation flux or the solar constant as considered to be constant at $200 \; W/m^2$ (see \cite{nasa} for benchmark values);
	\item No ventilation or airflow between zones is considered.
\end{enumerate}

Under these assumptions, \eqref{bilineardiscrete} simplifies to a linear, time-invariant dynamical system. The fact that ambient temperature and global radiation are assumed to be deterministic eliminate the input-disturbance bilinear terms. The resulting system is then denoted by
\begin{equation}
	x_{k+1} = A x_k + B_u u_k + B_\delta \delta_k, \label{eq:lti_sys}
\end{equation}
for appropriately defined matrices.

Out of the imposed assumptions, only the last one appears to be restrictive in practice, as it assumes absence of airflow or ventilation. The main focus of the paper, is to quantify the effect of using the algorithm in \cite{incscn} as an efficient, data driven manner to deal with uncertain occupancy. This requires the underlying optimization program to be convex, which in turn calls for linear dynamics; the latter is ensured under the imposed assumption. Current work concentrates towards relaxing the convexity assumption, thus allowing airflow or ventilation to be present, using the non-convex developments of \cite{waitjudge}.

\subsection{Energy Management Optimization}
\label{datadrivenoptimisation} 
We consider a finite-horizon energy management problem, where $M$ denotes the number of time steps. Let $X=[x_1^\top \ldots x_M^\top]^\top$ denotes a stacked vector including the states of all time instances, and define $U$ and $\delta$ similarly. 
Propagating the linear, time-invariant dynamics in \eqref{eq:lti_sys} by $M$ steps, we obtain the following compact representation of the systems temperature evolutions.
\begin{equation}
	X = F x_0 +G U + H \delta,
	\label{dynamiclong}
\end{equation}
where matrices $F,G$ and $H$ are of appropriate dimension; e.g., the reader is referred to \cite{simoneMPC} for dimensions. Note that $X$ depends on the history of inputs and disturbances that are included in $U$ and $\delta$, respectively, as well as on the initial state $x_0$, which is given.

We aim at minimizing the following objective function.
\begin{equation}
	J(U) = \mathbb{E} [ \,X^\top Q X + U^\top R U \,], \label{eq:obj}
\end{equation}
where we denote the objective function as being a function of $U$ only, as $X$ could be substituted by \eqref{dynamiclong} which is also a function of $U$ and the uncertainty vector $\delta$. 
We assume that $\delta$ is distributed according to some probability distribution $\mathbb{P}$ with support $\Delta$ (possibly unknown if only data are available), and denote by $\mathbb{E}$ the corresponding expectation operator.
The objective function in \eqref{eq:obj} involves the expected value of the sum of two components: one penalty on the state (temperature) and one on the input, where $Q \succeq 0$ and $R \succ 0$ are penalty matrices. In the numerical case study we only consider the second term, thus seeking a minimum effort sequence of actuation commands.

The minimization of $J(U)$ is subject to the following constraints.

\emph{Input constraints:}
The following constraints on the elements of the input vector $U$ are considered. For all $k$,
\begin{enumerate}
	\item $u_{1,k} \in (0,90)\%$: effective reduction of solar gain in the form of percentage for window blinds;
	\item $u_{2,k} \leq 1 kW/m^2$: upper limit on the heating system;
	\item $u_{3,k} \leq 1 kW/m^2$: upper limit on the cooling system.
\end{enumerate}  

\emph{State constraints:}
We differentiate between different seasons, namely, summer and winter, in line with the developments in \cite{Season}. This distinction allows us to model state constraints by means of single-sided inequality constraints; in the opposite case we would have double-sided inequalities; the latter may lead to feasibility issues, but this is outside the scope of this paper. We then have the following constraints on the temperature profiles in $X$ that encode comfort, which by means of \eqref{dynamiclong}, result in constraints on $U$: 
\begin{align}
	&\text{Summer day:} \nonumber \\
	&\; \mathbb{P} \{ \delta \in \Delta:~ F x_0+ GU+H\delta \leq T_{\max}\} \geq 1-\epsilon, 
	\label{constraint1} \\
	&\text{Winter day:} \nonumber \\
	& \; \mathbb{P} \{ \delta \in \Delta:~ F x_0+ GU+H\delta \geq T_{\min}\} \geq 1-\epsilon,
	\label{constraint2}
\end{align}
where $\epsilon \in (0,1)$.
Constraints \eqref{constraint1}, \eqref{constraint2}, are chance constraints, i.e., the guarantee that an upper ($T_{\max}$) and a lower ($T_{\min}$) temperature limit, respectively, is achieved, with probability at least $1-\epsilon$.

Note that we use an upper limit during summer because there exists a continuous heat flux from the surroundings,  and a lower limit for winter time as there exist a continuous heat flux to surroundings. For the considered case study, in winter time occupancy level provides enough heating for the temperatures to stay consistently above the lower limit; to highlight all features of the proposed algorithm we select a summer day for our numerical investigations.

By means of \eqref{dynamiclong}, we can collectively represent all input and state constraints outlines above as constraints involving only $U$ and $\delta$. To facilitate the algorithmic developments of the following section, the energy management problem under consideration is of the form
\begin{align}
\min_{U \in \mathbb{R}^{nM}}~ &J(U) \label{eq:comp_opt} \\
\text{subject to}~ & \mathbb{P} \{ \delta \in \Delta:~ g(U,\delta) \leq 0\} \geq 1-\epsilon, \label{eq:comp_con}
\end{align}
where $g$ is a scalar valued function (corresponding to the maximum among all constraints imposed on $U$), convex with respect to $U$ for any fixed $\delta$; the dependency on $\delta$ could be arbitrary. Note that in our formulation we have a total of $nM=3M$ decision variables, i.e., three scalar inputs times the number of time-steps $M$.

\section{Scenario approach implementation}
\label{algorithm}
\subsection{Standard scenario approach}
\label{oneshot}
The scenario approach provides a way to obtain a solution that is feasible for the chance constraint in \eqref{eq:comp_con} with certain confidence, without requiring knowledge of $\mathbb{P}$ and/or $\Delta$. To this end, associate with \eqref{eq:comp_opt}-\eqref{eq:comp_con} a scenario program, which involves replacing the chance constraint with $N$ constraints, each of them corresponding to realization/scenario of the uncertain occupancy level $\delta$, extracted in an i.i.d. (independent and identically distributed) fashion. This is given by
\begin{align}
\min_{U \in \mathbb{R}^{nM}}~ &J(U) \label{eq:comp_opt_sa} \\
\text{subject to}~& g(U,\delta^i) \leq 0, \text{ for all } i=1,\ldots,N. \label{eq:comp_con_sa}
\end{align}
We assume that \eqref{eq:comp_opt_sa}-\eqref{eq:comp_con_sa} is feasible for any multi-scenario extraction, and admits a unique solution, denoted by $U^*$. In case of multiple solutions,
 a convex tie-break rule could be adopted to single out a particular minimizer. The feasibility assumption can be relaxed as shown in \cite{Nsufficient}.

Under the feasibility and uniqueness requirements, the standard scenario approach \cite{scenario2}, shows that
for a given $\beta, \epsilon \in (0,1)$, if $N$ is chosen so that it satisfies
\begin{equation}
\sum_{j=0}^{nM-1} {N \choose j} \epsilon^j (1-\epsilon)^{N-j} \leq \beta, \label{eq:N}
\end{equation}
then
\begin{align}
	\mathbb{P}^N &\{(\delta^1,\ldots,\delta^N) \in \Delta^N:~  \nonumber \\
	&\mathbb{P} \{ \delta \in \Delta:~ g(U^*,\delta) \leq 0\} \geq 1-\epsilon \}
	\geq 1-\beta,
	\label{minN}
\end{align} 
where $\mathbb{P}^N$ denotes the product probability measure, as $U^*$ depends on all extracted scenarios used to solve \eqref{eq:comp_opt_sa}-\eqref{eq:comp_con_sa}, hence it is a random variable in $\Delta^m$.
In other words, with confidence at least $1-\beta$, the optimal solution $U^*$ of \eqref{eq:comp_opt_sa}-\eqref{eq:comp_con_sa} is feasible for the chance constraint in \eqref{eq:comp_con}. Note that the only structural characteristic appearing in the confidence is the number of decision variables $3M$.

We can obtain a sufficient, albeit explicit, condition for the number of scenarios $N$ that need to be extracted so that 
\eqref{eq:N} is satisfied. This is given by \cite{introbook}, \cite{simoneMPC},
\begin{equation}
	N \geq \frac{2}{\epsilon}\big(\ln \frac{1}{\beta} + nM\big).
	\label{scenarionumber}
\end{equation}

\subsection{Incremental scenario approach}
\label{incremental}
The incremental approach \cite{incscn}
leverages on the fact that the sample size proposed by the standard scenario approach is tight only for a certain class of programs (termed fully supported in \cite{scenario2}), and for problems typically encountered in applications may result in conservative performance.
The incremental scheme gradually introduces additional scenarios to a given initial set of scenarios and is guaranteed to result in a performance no worse than the standard scenario approach, thus reducing both the level of conservatism of the resulting solution and the sample size required. The general structure of the incremental scenario approach scheme is presented in Algorithm \ref{SCI}.

\begin{algorithm}
    \caption{Incremental scenario approach}
    \label{SCI}	
\begin{algorithmic}[1]
\STATE $j =  0$ , $N_{-1} = 0$ ,
 \WHILE{$S^*_j>j$}
 \STATE Add $(N_{j} - N_{j-1})$ additional scenarios
 \STATE Compute $U^*_j$ by solving \eqref{eq:comp_opt_sa}-\eqref{eq:comp_con_sa} with $N=N_j$
 \STATE Determine the number of support constraints $S^*_j$
 \IF{$S^*_j <j$} 
 \STATE Break and return $U^*=U^*_j$
 \ELSE
 \STATE $j \leftarrow j+1$
 \ENDIF
 \ENDWHILE
 \end{algorithmic}
\end{algorithm}

For each $j$ let $U_j^*$ denote the optimal solution of \eqref{eq:comp_opt_sa}-\eqref{eq:comp_con_sa} with $N=N_j$ (step 4).
Variable $S^*_j$ denotes the number of support constraints (a constraint is of support if its removal results in a different optimal solution \cite{introbook}) of $U_j^*$. It can be determined by means of an iterative process, where constraints are removed one by one, checking whether their removal resulted in a change in the optimal solution. For non-degenerate convex programs (see \cite{scenario2} for a formal definition of non-degeneracy), this is equivalent to determining the number of constraints active at the optimal solution; one way of achieving this is by considering the constraints corresponding to non-zero dual variables.

If at a given iteration $j$, $S_j^* < j$, then the algorithm terminates returning $U_j^*$ (step 7). This termination condition involves comparing the estimate of the support constraints with the iteration counter; this relationship between iterations and support constraints is due to the fact that the algorithm is guaranteed to terminate after at most $nM$ iterations, as the number of support constraints for convex programs can be no greater than the number of decision variables \cite{introbook}.
Due to this feature, the incremental scenario approach is guaranteed to require a number of scenarios no worse than the standard scenario approach.

In Theorem 1 in \cite{incscn}, it is shown that for non-degenerate problem instances, if the number of scenarios $N_0,N_1,\ldots,N_{nM}$ (recall that at most $nM$ algorithm iterations are required) is selected as shown in the sequel, then for a given $\beta, \epsilon \in (0,1)$,
\begin{align}
	\mathbb{P}^{nM} &\{(\delta^1,\ldots,\delta^N) \in \Delta^N:~  \nonumber \\
	&\mathbb{P} \{ \delta \in \Delta:~ g(U^*,\delta) \leq 0\} \geq 1-\epsilon \}
	&\geq 1-\beta,
	\label{minN}
\end{align}
where $U^*$ denotes the solution returned upon termination of Algorithm \ref{SCI}.
The number of scenarios required at each iteration $j=0,1,\ldots,nM$ is given by
\begin{align}
	&N_j = \min \bigg \{N : N\geq M_j \text{ and } \label{eq:N_j} \\
	&{N \choose j} (1-\epsilon)^{N-j} \leq \frac{\beta}{(d+1)(M_j+1)}\sum_{m=j}^{M_j} {m \choose j} (1-\epsilon)^{m-j} \bigg \}, \nonumber
\end{align}
where $M_j$ is given by 
\begin{equation}
	\label{Mjlong}
	M_j = \min \big \{ N : \sum_{\ell=0}^{j-1} {N \choose \ell } \epsilon^\ell (1-\epsilon)^{N-\ell} \leq \beta \big \},
\end{equation}

We now provide sufficient conditions for the satisfaction of the conditions in \eqref{Mjlong} and \eqref{eq:N_j}, to determine the number of scenarios $M_j$ and $N_j$, respectively. Due to the equivalence between \eqref{Mjlong} and \eqref{eq:N}, similarly to Section \ref{oneshot}, by \cite{introbook}, \cite{simoneMPC}, we obtain
\begin{equation}
	M_j = \frac{2}{\epsilon}\big(\ln \frac{1}{\beta} + j -1\big).
	\label{MJ}
\end{equation}
To obtain an explicit relation for $N_j$, set 
\begin{equation}
	\beta_j = \frac{\beta}{(d+1)(M_j+1)}\sum_{m=j}^{M_j} {m \choose j} (1-\epsilon)^{m-j}.
\end{equation}
We then seek a value for $N_j$ that satisfies ${N \choose j} (1-\epsilon)^{N-j} \leq \beta_j$. By Corollary 1 in \cite{Nsufficient},
we obtain 
\begin{equation}
	N_j = \big(\frac{2}{\epsilon} \ln \frac{1}{\beta_j}+2j+\frac{2j}{\epsilon}\ln \frac{2}{\epsilon}\big).
	\label{Nj}
\end{equation}

\section{Case study}
\label{simulation}
\subsection{Simulation set-up}
To quantify the efficacy of the proposed incremental scenario approach of Section \ref{incremental}, we compare it against a deterministic variant, where a nominal/forecasted occupancy level is considered, and against the standard scenario approach implementation of Section \ref{oneshot}. 

To this end, we revisit the energy management problem of Section \ref{secII}, considering \eqref{eq:obj} with $Q=0$ and $R=I$, thus resulting in minimizing $J(U) = U^T R U = U^T U$. Assuming a summer day, this is subject to \eqref{constraint1}, where we consider $T_{\max} = \ang{24} C$.
Moreover, the constraints on $U$ defined in Section \ref{secII} are also imposed. A time horizon of 12 hours with granularity of 15 minutes is considered, resulting in $M = 48$ time-steps.

We considered that the occupancy level follows a discrete Poisson distribution with a mean equal to $3$. All scenarios, for the standard and the incremental scenario approach, as well as the ones used for validation purposes, were generated in an i.i.d. fashion from the same distribution. For the deterministic analysis we considered occupancy to be the empirical expectation from a certain number of scenarios extracted from the aforementioned distribution.

\begin{table}[t]
\caption{Comparison between the deterministic, the standard, and the incremental scenario approach.}
\label{tableimportant}
\begin{tabular}{@{}c|cccc@{}}
\toprule
Algorithm                                                                   & \begin{tabular}[c]{@{}c@{}}Number of\\  Scenarios\end{tabular} & \begin{tabular}[c]{@{}c@{}}Cost ($\frac{W^2}{m^4}$)\end{tabular} & \begin{tabular}[c]{@{}c@{}}Theoretical\\  Risk \end{tabular} & \begin{tabular}[c]{@{}c@{}}Empirical\\  Risk\end{tabular} \\ \midrule
Deterministic                                                               & 1                                                              & $2.55\times 10^{4}$                                                & N/A                                                                       & $85.27 \%$                                                                    \\ \hline \\
\begin{tabular}[c]{@{}c@{}}Standard \\ Scenario \\ Approach\end{tabular}    & 3065                                                           & $2.36\times 10^{5}$                                                & $10\%$                                                                    & $0 \%$                                                                        \\ \hline \\
\begin{tabular}[c]{@{}c@{}}Incremental\\  Scenario \\ Approach\end{tabular} & 376                                                            & $1.94\times 10^{5}$                                                & $10\%$                                                                    & $5.1 \%$                                                                      \\ \bottomrule
\end{tabular}
\end{table}

\subsection{Simulation results}
We consider three algorithmic alternatives, namely, the deterministic approach, the standard scenario approach and the incremental one. We compare them in terms of the number of scenarios required by each alternative, resulting cost of the optimization program, theoretical risk (constraint violation level $\epsilon$) level, and empirical risk. The latter is determined by calculating the number of violations encountered by the solution returned by each approach over 3000 validation scenarios.
 
The comparison outcomes are summarised in Table \ref{tableimportant}.
The following observations are in order.\\
1) The cost incurred in the deterministic approach is significantly lower, however, this comes at the expense of not being robust with respect to uncertainty in the occupancy level. This can be witnessed by the high empirical risk, i.e., the empirical frequency of constraint violation against the 3000 validation scenarios. 
2) The standard scenario approach requires 3065, as opposed to 376 scenarios required by the incremental algorithm. Within a data driven context this is a desirable feature, in particular if data resources are limited.\\
3) As a result of the lower number of scenarios required for the same theoretical risk level $\epsilon = 10 \%$, the incremental algorithm leads to a less conservative behaviour. This is witnessed by the lower cost, as well as by the higher empirical risk ($5\%$ as opposed to $0\%$), which is closer to the theoretical value.

\begin{figure}[t]
	\centering
	\includegraphics[width = 9.3 cm]{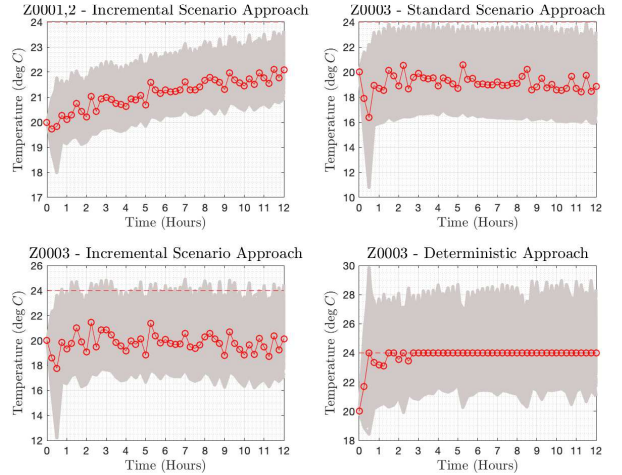}
	\caption{Temperature profile for zones Z0001 and Z0002 for the incremental scenario approach (they coincide as a result of the symmetry of the zones). Temperature profile for zone Z0003 using the incremental (lower left panel), the standard scenario approach (upper right panel), and the deterministic approach (lower right panel).}
	\label{tempprofile}
\end{figure}

Figure \ref{tempprofile} shows the temperature profiles of zone 3 (Z0003) generated by the three algorithms. The grey shadow corresponds to the span of the system trajectories when the optimal solution of each algorithmic alternative is monitored for the 3000 validation scenarios. The dashed red line is the upper bound for temperature during the summer. The red trajectory corresponds to the response of the system in the deterministic case, when occupancy is considered to be equal to the forecasted value. For completeness, the temperature profile of zones 1 (Z0001) and 2 (Z0002) is also illustrated for the incremental algorithm. 

Note that in the deterministic case (lower right panel of Figure \ref{tempprofile}), the vast majority of the validation scenarios leads to violation of the temperature limit. At the other extreme, no violation is encountered for the standard scenario approach (upper right panel of Figure \ref{tempprofile}). The incremental scenario approach results in a moderate number of violations, close to the theoretical acceptable risk level (lower left panel of Figure \ref{tempprofile}). 

\begin{figure}[t]
	\centering
	\includegraphics[width = 8.4 cm]{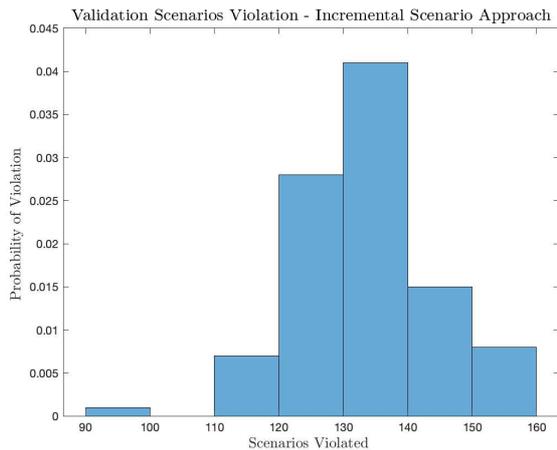}
	\caption{Histogram showing the empirical probability of constraint violation for 100 sets, each with 3000 scenarios. The empirical risk was calculated as the fraction out of 3000 validation scenarios that were violated by the solution returned by the incremental scenario approach algorithm.}
	\label{histogram}
\end{figure}

The probability of violation is itself a random variable, hence different sets of validation scenarios will result in a different empirical risks. We thus generated 100 sets, each with 3000 validation scenarios, and for each of them calculated the empirical risk. The resulting distribution is shown in Figure \ref{histogram}.
It must be noted that in all 100 runs of the validation process, not more than 160 scenarios were violated within each set of 3000 scenarios which is below the $10 \%$ theoretical risk level.

\section{Concluding remarks} \label{sec:concl}
We applied an incremental scenario approach algorithm recently introduced in \cite{incscn} to the problem of building energy management. To quantify the trade-off between performance and robustness we provide a detailed numerical analysis on a stylized building and compare it with respect to a deterministic design and the standard scenario approach implementation. 

Current work concentrates towards relaxing the convexity assumption, thus allowing airflow or ventilation to be present. To this end, we aim at employing the non-convex scenario approach developments of \cite{waitjudge}. Moreover, we aim at replacing the open-loop sequence of decisions with affine feedback policies with respect to uncertainty \cite{goulart}, and quantify the potential improvement in terms of performance. 

\bibliographystyle{ifacconf}
\bibliography{main.bib}        

\begin{thebibliography}{24}
\providecommand{\natexlab}[1]{#1}
\providecommand{\url}[1]{\texttt{#1}}
\providecommand{\urlprefix}{URL }
\expandafter\ifx\csname urlstyle\endcsname\relax
  \providecommand{\doi}[1]{doi:\discretionary{}{}{}#1}\else
  \providecommand{\doi}{doi:\discretionary{}{}{}\begingroup
  \urlstyle{rm}\Url}\fi

\bibitem[{{Bellushi} et~al.(2019){Bellushi}, {Falsone}, {Ioli}, {Margellos},
  {Garatti}, and {Prandini}}]{Prandini3}
{Bellushi}, F., {Falsone}, A., {Ioli}, D., {Margellos}, K., {Garatti}, S., and
  {Prandini}, M. (2019).
\newblock Distributed optimization for structured programs and its application
  to resource sharing in building district cooling.
\newblock In \emph{preprint submitted to Journal of Process Control}, 1--37.

\bibitem[{{Calafiore} and {Campi}(2006)}]{Nsufficient}
{Calafiore}, G.C. and {Campi}, M.C. (2006).
\newblock The scenario approach to robust control design.
\newblock \emph{IEEE Transactions on Automatic Control}, 51(5), 742--753.

\bibitem[{Campi and Garatti(2008)}]{scenario2}
Campi, M. and Garatti, S. (2008).
\newblock The exact feasibility of randomized solutions of uncertain convex
  programs.
\newblock \emph{SIAM Journal on Optimization}, 19(3), 1211--1230.
\newblock \doi{10.1137/07069821X}.

\bibitem[{Campi and Garatti(2018{\natexlab{a}})}]{introbook}
Campi, M. and Garatti, S. (2018{\natexlab{a}}).
\newblock \emph{Introduction to the Scenario Approach}.
\newblock \doi{10.1137/1.9781611975444}.

\bibitem[{Campi and Garatti(2018{\natexlab{b}})}]{waitjudge}
Campi, M. and Garatti, S. (2018{\natexlab{b}}).
\newblock Wait-and-judge scenario optimization.
\newblock \emph{Mathematical Programming}, 167(1), 155--189.

\bibitem[{Campi et~al.(2009)Campi, Garatti, and Prandini}]{MPCcontrol}
Campi, M.C., Garatti, S., and Prandini, M. (2009).
\newblock The scenario approach for systems and control design.
\newblock \emph{Annual Reviews in Control}, 33(2), 149 -- 157.
\newblock \doi{https://doi.org/10.1016/j.arcontrol.2009.07.001}.

\bibitem[{Campi et~al.(2019)Campi, Garatti, and Prandini}]{simoneMPC}
Campi, M.C., Garatti, S., and Prandini, M. (2019).
\newblock \emph{Scenario Optimization for MPC}, 445--463.
\newblock Springer International Publishing.

\bibitem[{{Causevich} et~al.(2018){Causevich}, {Falsone}, {Ioli}, and
  {Prandini}}]{Prandini1}
{Causevich}, V., {Falsone}, A., {Ioli}, D., and {Prandini}, M. (2018).
\newblock Energy management in a multi-building set-up via distributed
  stochastic optimisation.
\newblock In \emph{American Control Conference}, 1--6.

\bibitem[{{Chen} et~al.(2013){Chen}, {Wang}, {Heo}, and
  {Kishore}}]{MPCbuilding}
{Chen}, C., {Wang}, J., {Heo}, Y., and {Kishore}, S. (2013).
\newblock Mpc-based appliance scheduling for residential building energy
  management controller.
\newblock \emph{IEEE Transactions on Smart Grid}, 4(3), 1401--1410.
\newblock \doi{10.1109/TSG.2013.2265239}.

\bibitem[{{Garatti} and {Campi}(2019)}]{incscn}
{Garatti}, S. and {Campi}, M.C. (2019).
\newblock Complexity-based modulation of the data-set in scenario optimization.
\newblock In \emph{2019 18th European Control Conference (ECC)}, 1386--1391.

\bibitem[{Goulart et~al.(2006)Goulart, Kerrigan, and Maciejowski}]{goulart}
Goulart, P.J., Kerrigan, E.C., and Maciejowski, J.M. (2006).
\newblock Optimization over state feedback policies for robust control with
  constraints.
\newblock \emph{Automatica}, 42(4), 523 -- 533.
\newblock \doi{https://doi.org/10.1016/j.automatica.2005.08.023}.

\bibitem[{Hussain et~al.(2017)Hussain, Bui, Kim, Im, and Lee}]{Season}
Hussain, A., Bui, V.H., Kim, H.M., Im, Y.H., and Lee, J.Y. (2017).
\newblock Optimal energy management of combined cooling, heat and power in
  different demand type buildings considering seasonal demand variations.
\newblock \emph{Energies}, 10(6), 789.

\bibitem[{{Ioli} et~al.(2016){Ioli}, {Falsone}, and {Prandini}}]{Prandini2}
{Ioli}, D., {Falsone}, A., and {Prandini}, M. (2016).
\newblock Energy management of a building cooling system with thermal storage:
  a randomized solution with feedforward disturbance compensation.
\newblock In \emph{American Control Conference}, 2378--5861.

\bibitem[{Jacomino and Le(2012)}]{robust1}
Jacomino, M. and Le, M.H. (2012).
\newblock Robust energy planning in buildings with energy and comfort costs.
\newblock \emph{4OR}, 10(1), 81--103.

\bibitem[{Lehmann et~al.(2013)Lehmann, Gyalistras, Gwerder, Wirth, and
  Carl}]{swissoffice}
Lehmann, B., Gyalistras, D., Gwerder, M., Wirth, K., and Carl, S. (2013).
\newblock Intermediate complexity model for model predictive control of
  integrated room automation.
\newblock \emph{Energy and Buildings}, 58, 250 -- 262.
\newblock \doi{https://doi.org/10.1016/j.enbuild.2012.12.007}.

\bibitem[{NASA(2019)}]{nasa}
NASA (2019).
\newblock Net radiation data by nasa.

\bibitem[{{Oldewurtel} et~al.(2012){Oldewurtel}, {Parisio}, {Jones},
  {Gyalistras}, {Gwerder}, {Stauch}, {Lehmann}, and {Morari}}]{Frauke1}
{Oldewurtel}, F., {Parisio}, A., {Jones}, C., {Gyalistras}, D., {Gwerder}, M.,
  {Stauch}, V., {Lehmann}, B., and {Morari}, M. (2012).
\newblock Use of model predictive control and weather forecasts for energy
  efficient building climate control.
\newblock In \emph{Energy and Buildings}, 45, 15--27.

\bibitem[{{Oldewurtel} et~al.(2010){Oldewurtel}, {Parisio}, {Jones}, {Morari},
  {Gyalistras}, {Gwerder}, {Stauch}, {Lehmann}, and {Wirth}}]{Frauke2}
{Oldewurtel}, F., {Parisio}, A., {Jones}, C., {Morari}, M., {Gyalistras}, D.,
  {Gwerder}, M., {Stauch}, V., {Lehmann}, B., and {Wirth}, K. (2010).
\newblock Energy efficient building climate control using stochastic model
  predictive control and weather predictions.
\newblock In \emph{American Control Conference}, 5100--5105.

\bibitem[{Palmer and Cooper(2013)}]{department1}
Palmer, J. and Cooper, I. (2013).
\newblock United kingdom housing energy fact file.

\bibitem[{Saha et~al.(2015)Saha, Kuzlu, Pipattanasomporn, Rahman, Elma,
  Selamogullari, Uzunoglu, and Yagcitekin}]{robust2}
Saha, A., Kuzlu, M., Pipattanasomporn, M., Rahman, S., Elma, O., Selamogullari,
  U.S., Uzunoglu, M., and Yagcitekin, B. (2015).
\newblock A robust building energy management algorithm validated in a smart
  house environment.
\newblock \emph{Intelligent Industrial Systems}, 1(2), 63--174.

\bibitem[{{Sturzenegger} et~al.(2016){Sturzenegger}, {Gyalistras}, {Morari},
  and {Smith}}]{MPC1}
{Sturzenegger}, D., {Gyalistras}, D., {Morari}, M., and {Smith}, R.S. (2016).
\newblock Model predictive climate control of a swiss office building:
  Implementation, results, and cost–benefit analysis.
\newblock \emph{IEEE Transactions on Control Systems Technology}, 24(1), 1--12.
\newblock \doi{10.1109/TCST.2015.2415411}.

\bibitem[{{Sturzenegger} et~al.(2014){Sturzenegger}, {Gyalistras}, {Semeraro},
  {Morari}, and {Smith}}]{ETH1}
{Sturzenegger}, D., {Gyalistras}, D., {Semeraro}, V., {Morari}, M., and
  {Smith}, R.S. (2014).
\newblock Brcm matlab toolbox: Model generation for model predictive building
  control.
\newblock In \emph{2014 American Control Conference}, 1063--1069.

\bibitem[{Sturzenegger et~al.(2012)Sturzenegger, Gyalistras, Morari, and
  Smith}]{ETH2}
Sturzenegger, D., Gyalistras, D., Morari, M., and Smith, R.S. (2012).
\newblock Semi-automated modular modeling of buildings for model predictive
  control.
\newblock BuildSys '12, 99--106. ACM.
\newblock \doi{10.1145/2422531.2422550}.

\bibitem[{Waters(2018)}]{department2}
Waters, L. (2018).
\newblock Energy consumption in the uk (ecuk) 1970 to 2018.

\end{thebibliography}
\end{document}